\documentclass[aps,prl,reprint,groupedaddress]{revtex4-2}
\usepackage{float}
\usepackage{amsmath}
\usepackage{amssymb}
\usepackage{graphicx} 
\usepackage{color}
\usepackage{braket}
\usepackage{enumitem}

\definecolor{linkcol}{RGB}{18, 94, 173}     
\definecolor{citecol}{RGB}{166, 45, 124}    
\definecolor{urlcol}{RGB}{0, 128, 112}      

\usepackage[
    colorlinks=true,
    linkcolor=linkcol,
    citecolor=citecol,
    urlcolor=urlcol,
]{hyperref}

\def\be{\begin{equation}}
\def\ee{\end{equation}}
\def\ba{\begin{align}}
\def\ea{\end{align}}
\def\bs{\begin{split}}
\def\es{\end{split}}
\def\bea{\begin{eqnarray}}
\def\eea{\end{eqnarray}}

\newcommand{\order}{\mathcal{O}}

\def\reals{\mathbb{R}}

\definecolor{darkred}{rgb}{0.8,0.1,0.1}

\renewcommand{\d}{\mathrm{d}}
\newcommand{\upd}{\mathrm{d}}
\newcommand{\e}{\mathrm{e}}
\begin{document}

\title{Extracting quantum field theory dynamics from an approximate ground state}

\author{Sophie Mutzel}
\email[]{sophie.mutzel@minesparis.psl.eu}
\author{Antoine Tilloy}
\email{antoine.tilloy@minesparis.psl.eu}

\affiliation{Laboratoire de Physique de l’École Normale Supérieure, Mines Paris, Inria, CNRS, ENS-PSL, Centre Automatique et Systèmes (CAS),
Sorbonne Université, PSL Research University, Paris, France}

\date{\today}

\begin{abstract}
    We develop a linear-programming method to extract dynamical information from static ground-state correlators in quantum field theory. We recast the Källén–Lehmann inversion as a convex optimization problem, in a spirit similar to the recent approach of Lawrence \href{https://arxiv.org/abs/2408.11766}{[\texttt{2408.11766}]}. This produces robust estimates of the smeared spectral density, the real-time propagator, and the mass gap directly from an approximate equal-time two-point function, and simultaneously yields an \emph{a posteriori} lower bound on the correlation-function error. We test the method on the $1+1$-dimensional $\phi^4$ model, using a variational approximation to the vacuum---relativistic continuous matrix product states---that provides accurate correlators in the continuum and thermodynamic limits. The resulting mass gaps agree with renormalized Hamiltonian truncation and Borel-resummed perturbation theory across a wide range of couplings, demonstrating that accurate dynamical data can be recovered from a single equal-time slice.
\end{abstract}

\maketitle

\noindent \textit{Introduction} --
In many non-perturbative approaches to relativistic quantum field theory (QFT), one has easy access only to \emph{Euclidean} correlation functions of local operators. In principle, knowing all these correlation functions fully determines the theory, including its real-time dynamics, using the Osterwalder–Schrader theorem~\cite{Osterwalder:1974tc,Wightman}. It may however be difficult to extract this data in practice.

With variational methods, like tensor network states~\cite{schollwock2011mpsreview,cirac2009tnreview}, one usually has access to an even more restricted set of correlation functions. The idea of variational methods is to minimize the energy over a finite-dimensional submanifold of the continuum Hilbert space, which provides a good approximation  $|\psi_0\rangle$ of the interacting vacuum $\ket{\Omega}$. Given this state, one gets immediate access to \emph{static} (or equal-time) correlation functions,
\begin{equation}\label{eq:static_correlations}
	C(x_1,\ldots,x_n)=\langle \psi_0 |\phi(0,x_1)\ldots \phi(0,x_n) |\psi_0\rangle \, ,
\end{equation}
which are often extremely accurate. This naturally raises the question of how much \emph{dynamical} data one can extract, in principle and in practice, from such a convenient subset of \emph{static} Euclidean correlation functions.

Applying the variational principle to a relativistic QFT and directly in the continuum is more demanding than for a lattice model~\cite{Feynman}. However, once the corresponding challenges are addressed, we get far more information from the variational ground state. First, being in the continuum allows to evaluate the static correlation functions \eqref{eq:static_correlations} at arbitrary points. For example, one may span a wide range of scales by probing logarithmically spaced positions $x$, which would be exponentially expensive on the lattice. Second, because of Euclidean invariance in imaginary time, equal-time correlation functions along the spatial direction are \emph{equal} to correlation functions along the Euclidean time direction. For example $
	\braket{\phi(\tau,0)\phi(0,0)}_E = \braket{\phi(0,\tau)\phi(0,0)}_E
$,
which implies, in particular, that the mass gap is \emph{exactly} encoded in the \emph{spatial} decay of vacuum correlation functions.

Focusing on two-point functions, there is a direct relation between the two-point correlation function and the spectrum of the theory via the Källén-Lehmann spectral representation \cite{Lehmann:1954xi,Kallen:1952zz,Kallen:1972pu}. For a scalar field in Euclidean time it takes the form:
\begin{equation}\label{eq:KL}
	\langle\phi(x) \phi(y)\rangle_E=\int_0^{\infty} \d s \, \rho(s)\, G_E(x-y; s) \;,
\end{equation}
where the \emph{spectral density} $\rho(s)$ is a \emph{positive} function and $G_E$ is the Euclidean propagator of a free scalar particle with mass $\sqrt{s}$. In $1+1$ dimensions, it is
\begin{equation}\label{eq:propagator}
	G_E(x,s) := \int_{\mathbb{R}^2} \frac{\upd^2p}{(2\pi)^2}  \frac{\e^{- ip\cdot x}}{p^2 + s} = \frac{1}{2\pi} K_0(s\, \|x\|)\;,
\end{equation}
where $K_0$ is the modified Bessel function of the second kind. For a massive scalar theory with repulsive interactions $\rho(s)$ takes the form
\begin{equation}\label{eq:rho_scalar_1d}
	\rho(s) = Z_\phi \delta(M^2-s) + \tilde{\rho}(s)\theta(s-s_{\mathrm{th}})\;,
\end{equation}
\textit{i.e.}~a one-particle state with mass $M$ (the gap of the theory) and a regular continuum part typically starting at the threshold for pair creation $s_{\mathrm{th}} = 4M^2$, where particles can go on-shell. The spectral density also gives access to real-time data, because eq.~\eqref{eq:KL} can be continued into:
\begin{equation}\label{eq:KL_realtime}
	\langle\Omega| T\{\phi(x) \phi(y)\}|\Omega\rangle=\int_0^{\infty} \d s \, \rho(s) D_F(x-y ; s) \;,
\end{equation}
where $D_F$ is the Feynman propagator for a free particle of mass $\sqrt{s}$. Using the optical theorem, $\rho(s)$ can be directly related to inclusive cross sections. 

\begin{figure}
	\centering
	\includegraphics[width=0.5\linewidth]{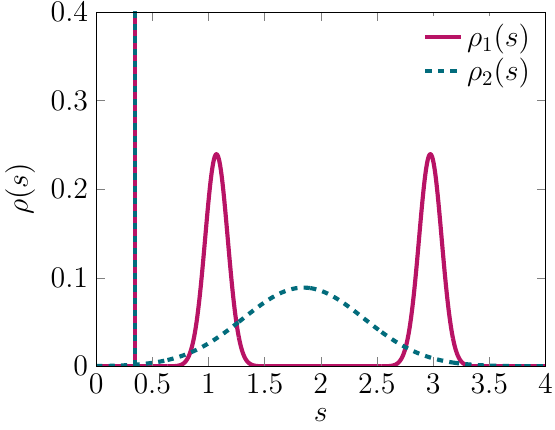}\includegraphics[width=0.5\linewidth]{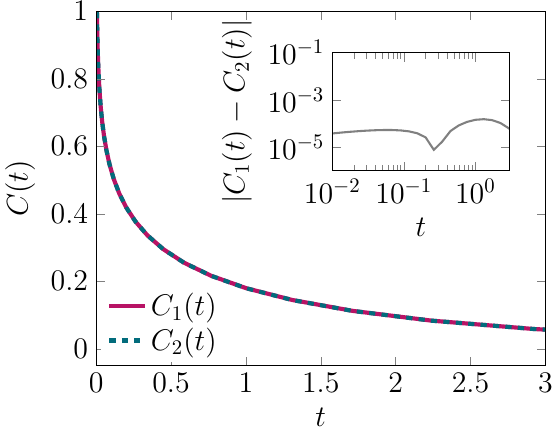}
	\caption{
		Left: Two spectral densities with identical one particle state but very different multiparticle continua. Right: Corresponding Euclidean two-point functions obtained via the Källén-Lehmann relation \eqref{eq:KL}. The absolute difference between both correlators is of the order $10^{-4}$.  }
	\label{fig:KL}
\end{figure}

We would thus obtain a wealth of dynamical data if we could invert \eqref{eq:KL}, and get the spectral density as a function of static correlation functions.
Unfortunately, it is well known that when the correlator is only known at discrete points and affected by numerical errors, this inversion is a mathematically ill-posed problem. Namely, infinitely many distinct spectral densities can reproduce the same Euclidean data within numerical accuracy, and their pointwise differences can be arbitrarily large. This is illustrated in Fig.~\ref{fig:KL}, where two spectral densities with identical one-particle masses but different multiparticle continua yield nearly indistinguishable two-point functions, differing by less than $10^{-4}$.

Several techniques have been developed to reconstruct $\rho$ approximately from lattice Monte-Carlo data. 
They include maximum-entropy and Bayesian reconstructions \cite{Jarrell:1996,Asakawa:2000tr,PhysRevLett.111.182003,PhysRevD.95.056016}, Padé–based analytic continuation \cite{Tripolt:2018xeo}, stochastic and neural-network approaches \cite{PhysRevB.57.10287,Buzzicotti:2023qdv,Wang:2021jou,PhysRevLett.124.056401,Horak:2021syv}, extensions of Backus–Gilbert \cite{Hansen:2019idp}, analytic continuation methods \cite{Bergamaschi:2023xzx} and Mellin-transform techniques for spectral densities \cite{Bruno:2024fqc}.
However, most of these approaches require either regularization or implicit model assumptions to stabilize the inversion. 
Recently, a linear programming approach has been proposed by Lawrence~\cite{Lawrence:2024hjm}. It allows to leverage the positivity of $\rho$ to get a stable and agnostic reconstruction. This method has shown promise on the lattice, but it is not tailored to the strengths of variational methods and of the continuum yet. 

In this letter, we introduce a linear-programming approach, inspired by \cite{Lawrence:2024hjm}, to extract precise spectral and real-time information directly from static correlators. The method requires only that the variational ansatz provide deterministic, continuum, thermodynamic-limit two-point functions at arbitrary separations, albeit with an intrinsic and unknowable variational error. Any variational scheme with these properties is suitable and we illustrate the approach using relativistic continuous matrix product states (RCMPS). A notable byproduct is an \emph{a posteriori} lower bound on the variational error itself.

\paragraph{Setup} --~~ Throughout this letter, we consider a simple self-interacting scalar field with Hamiltonian
\begin{equation}\label{eq:hamiltonian}
	H=\int_\mathbb{R} \d x : \frac{ \pi^2}{2}+\frac{(\nabla \phi)^2}{2}+\frac{m^2}{2} \phi^2+g \phi^4 : \; .
\end{equation}
This is a well-defined model, with energy density bounded below~\cite{glimm1968_phi4bound,federbush1969_phi4bound}, but that is not exactly solvable. The physics depends only on $g/m^2$ and so we can take $m=1$ without loss of generality. We stick to a quartic potential for simplicity but our analysis would work just as well for any other polynomial.

We approximate the ground state of \eqref{eq:hamiltonian} by minimizing the corresponding energy density over the manifold of RCMPS \cite{Tilloy:2021yre,Tilloy:2021hhb,Tilloy:2022kcn}. These states are parameterized by two complex matrices $Q,R$ of size $D\times D$, where $D$ is the \emph{bond dimension}, which controls their expressiveness. We present RCMPS in more detail in the End Matter. All that is needed for us now is that one can efficiently minimize the energy density over $Q,R$ with Riemannian optimization techniques, at a cost $\propto D^3$ per iteration. Although no rigorous convergence result is known, it is observed that local observables converge fast (possibly super-polynomially fast) as a function of $D$. Once an approximate ground state $|\psi_0\rangle = |Q,R\rangle$ is known, one can compute the two-point correlation function $C_{D}(x):=\langle Q,R|\phi(0,x) \phi(0,0) |Q,R\rangle $ at arbitrary points and in a numerically exact way. In what follows, we thus assume as starting point that we are given $C_{D}(x_k)$ for an arbitrary number of points.

\paragraph{Spectral density from convex optimization} --~~ 
The core idea of our method, following \cite{Lawrence:2024hjm}, is to transform the inversion problem \eqref{eq:KL} into a convex optimization problem. To this end, we ask: how large (or small) could the spectral density be at a certain point $s$, given that it has to be positive everywhere and yield the correlation function we computed via \eqref{eq:KL}?

The spectral density itself is a singular object generically -- eq. \eqref{eq:rho_scalar_1d} shows it typically contains a Dirac mass -- and so it is hopeless to bound it point-wise. Instead, we need to consider how large a certain test function integrated against it could be. We start by taking a Gaussian as test function, and seek to maximize or minimize:
\begin{equation}
\rho^\sigma(\mu^2)\equiv \frac{1}{\sqrt{2 \pi} \sigma} \int_{\mathbb{R}^{+}} \rho(s) \ \mathrm{e}^{-(\mu^2-s)^2 /\left(2 \sigma^2\right)} \mathrm{d} s \;,
\end{equation}
which is finite.

Including the general physical constraints on $\rho$, and the fact that it needs to reproduce the function $C_{D}(x)$ we computed, leads to the following constrained optimization problem over functions \begin{subequations}\label{eq:linprobinf}
\begin{align}
\max\limits_{\rho}/\min\limits_{\rho} \quad & \rho^\sigma(\mu^2) \label{eq:cost}\\ 
\text { under } 
& \cdot \rho(s) \geq 0 \quad \forall s \geq 0 \label{eq:pos}\\ 
& \cdot \int \mathrm{~d} s \ \rho(s)=1 \label{eq:int}\\ 
& \cdot C_{D}(x)=\int_{\mathbb{R}^{+}} \mathrm{d} s \ \rho(s) G_E(x;s) \quad \forall x \in \mathbb{R} \;. \label{eq:KLprob}
\end{align}
\end{subequations}
Importantly, it is a linear program: we optimize a linear function \eqref{eq:cost} of the variables under the linear inequalities \eqref{eq:pos} and under the linear equalities \eqref{eq:int} and \eqref{eq:KLprob}.

\begin{figure}
    \centering
    \includegraphics[width=1\linewidth]{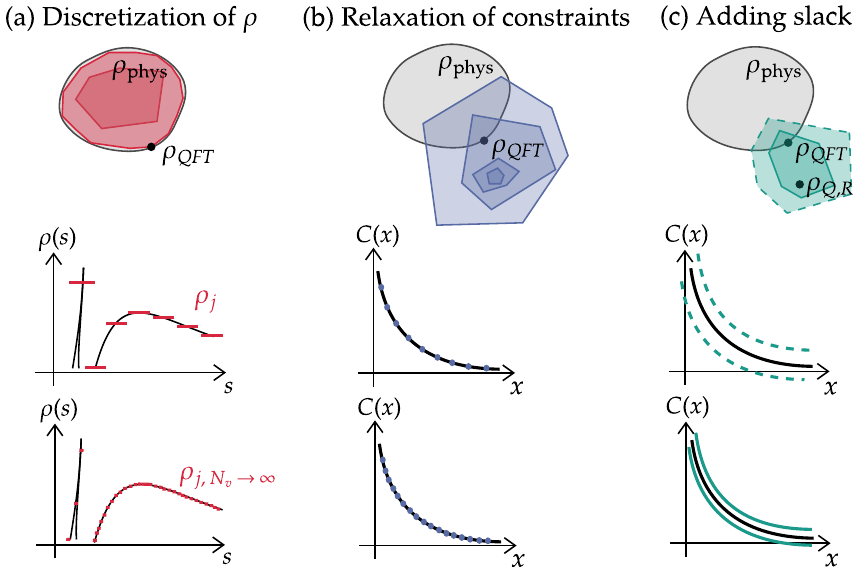}
    \caption{(a)-(c) Top: illustrative projection of convex space of physically allowed spectral densities for the various relaxations/approximations described in the main text. 
    Bottom: Relaxations on $C(x)$ or approximations on $\rho(s)$ of the problem \eqref{eq:LPreduced}.    
    (a) Approximation by discretizing in $s$, which can be controlled  by taking a large number of variables $N_v$ for discretization in $s$. 
    (b)  Relaxation of original problem \eqref{eq:linprobinf} by restricting to finite number of points on the correlator. 
    (c) Relaxation by allowing a certain slack $\delta C$ on correlator. }
    \label{fig:convex}
\end{figure}

However, the problem is infinite both in variables ($\rho(s)$) and constraints ($x\in \reals$). To render it numerically tractable, we both relax and discretize it as follows:
    \begin{enumerate}[label=(\roman*),nosep]
        \item we consider the constraint on the correlator only on a subset of points $x_k$;
        \item we allow a finite slack $\delta C$ between $C_{D}$ and the reconstructed correlation function, to account for the small systematic errors in our RCMPS results \footnote{We find that without this tolerance, no non-negative $\rho(s)$ would satisfy the constraints, meaning our RCMPS two-point function is inconsistent with a non-negative spectral function.};
        \item we discretize $s$ into equidistant grid points $s_j=(s_{\max}-s_{\min})/N_v$ and hence $\rho(s_j)=\rho_j$ into a finite number of variables, with $\Delta_j=s_{j+1}-s_j$ the corresponding integration weights.
    \end{enumerate}
Steps (i)–(ii) are true relaxations of the original problem \eqref{eq:linprobinf}: since the feasible region enlarges, both bounds loosen but remain rigorous. This is illustrated in Fig.~\ref{fig:convex}
(b)-(c). Step (iii) is an approximation, introducing a controllable numerical error, which we need to make sure is negligible. In practice, we use $N_v\sim 10^4$ grid points and verify stability under refinement, as represented in Fig.~\ref{fig:convex} (a). 
With these approximations and relaxations, we obtain a finite dimensional optimization problem
\renewcommand{\arraystretch}{1.5}
\begin{equation}
  \label{eq:LPreduced}
\begin{array}{cl}
\max\limits_{\rho_{j}}/\min\limits_{\rho_{j}} 
& \rho_{j}^\sigma(\mu^2) \\ 
\text { under } 
& \cdot \rho_{j} \geq 0\;, \quad  j=1,\ldots , N_v \\ 
& \cdot \sum_j \ \rho_{j} \Delta_{j} = 1 \\ 
& \cdot C_D(x_i) - \delta C \leq  \sum_j \Delta_{j} \ \rho_{j} G_E (x_i;s_{j}) \\
& \cdot  C_D(x_i) +  \delta C \geq \sum_j \Delta_{j} \ \rho_{j} G_E (x_i,s_j)\;, \\
&\text{for   } i= 1,\ldots , N_c\;, \\
 \end{array} 
\end{equation}
which can be efficiently solved using standard convex optimization techniques (e.g., the Simplex algorithm) \footnote{For the numerical solutions of the optimization problem presented in this letter, we used the Julia JUMP library \cite{DunningHuchetteLubin2017} to set up the linear program, and solved it using IBM ILOG CPLEX.}. 

This strategy can also be applied to bound real-time correlators directly. To this end, we can just replace the Gaussian smearing in \eqref{eq:LPreduced} with the appropriate weight. For instance, the retarded propagator is obtained by solving the same linear problem as before, but with the optimization of the linear cost function: 
\begin{equation}\label{eq:cost_realtime}
\max\limits_{\rho}/\min\limits_{\rho} \quad G_R(\rho  ; t,0) = \frac{1}{2} \int_0^\infty \d s\ \rho(s) J_0(\sqrt{s}t)
\end{equation}
with $J_0$ the Bessel function of the first kind \footnote{The Bessel function $J_0$ is unbounded in $s=0$. One may fear that it could make the maximization problem ill-defined. Fortunately, for a gapped theory, we know $\rho$ has no support near $s=0$ and can thus set $\rho(s\leq s_{\mathrm{reg}})=0$ for some sufficiently small $s_{\mathrm{reg}}$, \textit{e.g.} $s_{\mathrm{reg}}=M^2/10$ where $M$ is a crude approximation to the mass gap.}.
This problem can be relaxed and discretized as before.

\paragraph{Lower bounding the slack $\delta C$} --~~
If the systematic error $\delta C$ were known (and the discretization of $\rho(s)$ arbitrarily fine), the upper and lower bounds on the smeared spectral density or real-time correlation functions obtained by solving problem \eqref{eq:LPreduced} would be fully rigorous. However, this error is not knowable \emph{a priori}: while a variational method provides rigorous energy upper bounds, we merely know that other observables converge as the variational manifold covers the Hilbert space. We thus turn the problem around. Instead of taking the error $\delta C$ as input, we ask: How small can the error $\delta C$ be while keeping the correlation function data consistent with a physical spectral density $\rho \geq 0$? Interestingly, this gives us as a byproduct a non-trivial \emph{a posteriori} lower bound on RCMPS correlation function errors.

In practice, we considered a simple absolute error model, \textit{i.e.\!} $\delta C$ independent of $x$. This seems consistent with our RCMPS data, at least for $x$ not too large. Indeed, by comparing $C_{D}(x)$ at different bond dimensions $D$, we observed that the error is approximately constant for $x\lesssim \text{a few}\, M^{-1}$ where $M$ is the mass gap. This motivated us to take logarithmically spaced points $x_k$ in the interval $[10^{-5}, \text{a few}\, M^{-1}]$, where the constant error model seems reasonable, and we still capture the relevant physics. The results do not strongly depend on these bounds.

We now bootstrap $\rho^{\sigma}(\mu^2)$ (or $G_R(t)$) and the minimal error $\delta C$ together as follows: 
We start from a large constant $\delta C$, such that the linear problem \eqref{eq:LPreduced} is feasible ($C_{D}(x)\pm \delta C$ is consistent with a non-negative $\rho(s)$), and gradually increase the number $N_c$ of constraint points, until the upper and lower bounds on $\rho^\sigma(\mu^2)$ each converge as a function of $N_c$ (which happens for $N_c\sim 100$). We then reduce $\delta C$ until the problem becomes unfeasible, and locate the edge of feasibility by bisection. The solution defines both our best estimate of the smeared spectral density and the minimal RCMPS error.

\begin{figure}
    \centering
    \includegraphics[width=1\linewidth]{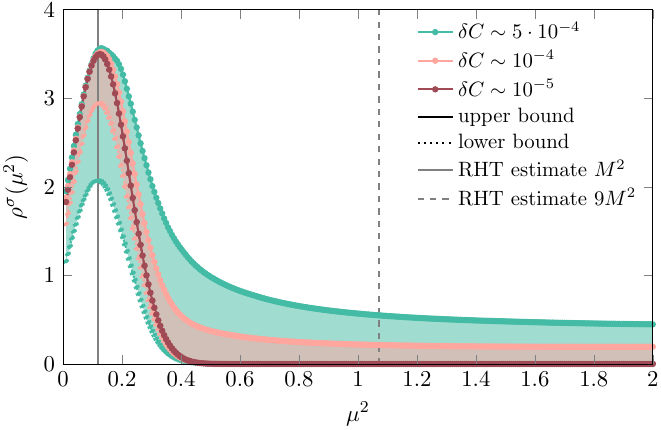}
    \caption{Smeared spectral densities ($\sigma^2=0.01$) obtained from the linear program \eqref{eq:LPreduced}  for $\phi^4$-theory at $g=2$. We use $N_c \sim 100$ constraints in the range  $x\in[10^{-5},3]$ calculated on an RCMPS ground-state approximation with bond dimension $D=44$. For comparison, we show the mass-gap estimate from renormalized Hamiltonian truncation \cite{Elias-Miro:2017xxf}, and the expected start of the multiparticle threshold at $9M^2$.} 
\label{fig:rho_smeared} \end{figure}

We illustrate this procedure in Fig.~\ref{fig:rho_smeared} for the $\phi^4$ model at coupling $g=2$. We show the upper and lower bounds on the smeared spectral density $\rho^{\sigma}(\mu^2)$, for various choices of $\delta C$.  Our best estimate for the smeared spectral density is the line where both the upper and lower bound coincide, which happens for $\delta C \sim 10^{-5}$. Below this threshold, no non-negative spectral density is consistent with the RCMPS correlator, and we thus know our RCMPS error must be larger. 

The resulting spectrum correctly reproduces the one-particle pole without imposing any prior structure. Its position (the mass gap), also agrees with renormalized Hamiltonian truncation results.

However, reconstructing fine local features of the spectrum, such as the precise shape of the multi-particle continuum, remains challenging. This behavior is expected from the numerical sensitivity, in Fig.~\ref{fig:KL}: resolving high-energy structures with small weight would require the Euclidean correlator to be known to machine precision or better. Attempting a perfect reconstruction of the full (smeared) spectrum is therefore overly ambitious. Instead, for large $s$, one should consider only robust global observables, obtained by integrating $\rho(s)$ on larger windows. This still provides access to nontrivial dynamical information, such as the wave-function renormalization constant $Z_\phi$, as we discuss in the End Matter. 
\begin{figure}
    \centering
\includegraphics[width=1\linewidth]{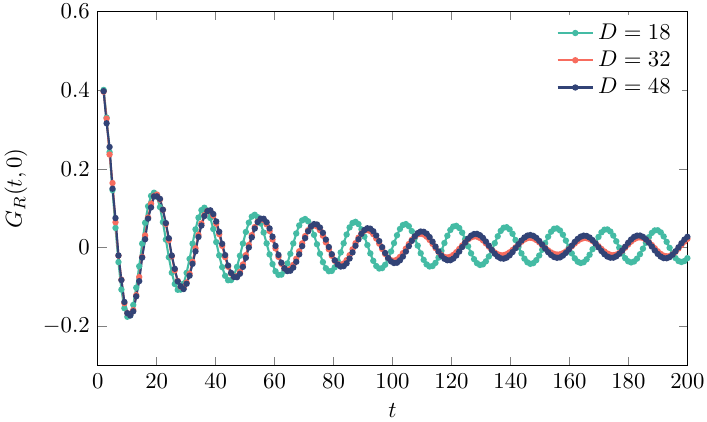}
\caption{Retarded propagator $G_R(t,0)$, see Eq.~\eqref{eq:cost_realtime}, obtained by numerically solving the linear program \eqref{eq:LPreduced} for $\phi^4$-theory at coupling $g=2$ and by bootstrapping the RCMPS slack $\delta C$, until the upper and lower bound coincide. We took $N_c\sim 100$ logaritmically spaced points for the correlation function in the range $x\in [10^{-3},3]$.\label{fig:real-time}}
\end{figure}

We also show in Fig.~\ref{fig:real-time} the estimates for the real-time evolution of the retarded propagator. While there is clearly a dephasing visible for $D=18$, larger values $D=32$ and $D=48$ can barely be distinguished until $t=200$. This suggests that accurate real-time dynamics can be recovered directly from static RCMPS data. Obtaining such results with conventional tensor-network approaches would require more work: one would need to evolve the ansatz in real-time with the time dependent variational principle (TDVP)~\cite{haegeman2011_original_tdvpMPS,vanderstraeten2019_tangentspace}, which would break translation invariance (something that is not yet doable efficiently with RCMPS).

\paragraph{Bootstrapping the mass gap} --~~
The physical mass $M$ is a particularly important quantity. In principle, we could infer it from the position of the (smeared) Dirac peak in Fig.~\ref{fig:rho_smeared}, but this is a fairly indirect approach. In our case, the form of the spectral density is partly known (see eq.~\eqref{eq:rho_scalar_1d}), which is something we can exploit. We may replace the variable $\rho(s)$ by $M$, $Z_\phi$, and the regular part $\tilde{\rho}(s)$ of the spectral density past threashold. In these new variables, the problem \eqref{eq:linprobinf} is a linear program, and thus efficiently solvable, only if $M$ is kept fixed.

Intuitively, for values of $M$ that are too far from the true value, the linear program \eqref{eq:linprobinf} has no feasible solution consistent with the RCMPS correlation function. We thus use a boostrap strategy: starting from a large slack $\delta C$, we scan over $M$ to find a range of values $M\in [\sqrt{s_1},\sqrt{s_2}]$ for which the constraints can be satisfied and the problem is feasible. We then decrease $\delta C$ until this interval collapses to a single value $M_\mathrm{opt}$, which we locate efficiently by bisection. This is our best estimate of the mass gap. As before, the value of $\delta C$ at the optimum provides a lower bound on the systematic correlation function error. 

\begin{figure}
    \centering
    \includegraphics[width=1\linewidth]{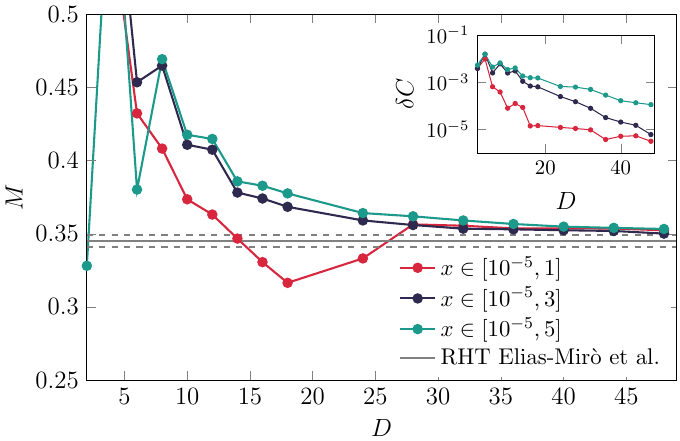} 
    \caption{       
Mass gap estimated with the bootstrap approach for $g=2$ using the
RCMPS value of the correlator in the domains $x\in [10^{-5},5]$, $x\in [10^{-5},3]$ and $x\in [10^{-5},1]$, as a function of the bond dimension. On the top right, we plot the corresponding output of the systematic error on RCMPS correlation functions. In gray, we also compare to the carefully extrapolated Renormalized Hamiltonian Truncation results of \cite{Elias-Miro:2017xxf}. The gray dashed lines correspond to a $2\sigma$ distance from the mean, obtained from the estimated systematics reported in \cite{Elias-Miro:2017xxf}.}
    \label{fig:bootstrap_g2}
\end{figure}

The resulting $M_\text{opt}$ as a function of $D$ for fixed $g=2$ are shown in Fig.~\ref{fig:bootstrap_g2}. For $D\gtrsim 28$, the estimates from different choices of constraint intervals collapse, and converge well to the renormalized Hamiltonian truncation values. Hence, as expected, our method is not really sensitive to the precise choice of interval as $D$ is increased and the RCMPS results become more accurate.

\begin{figure}
    \centering
    \includegraphics[width=1\linewidth]{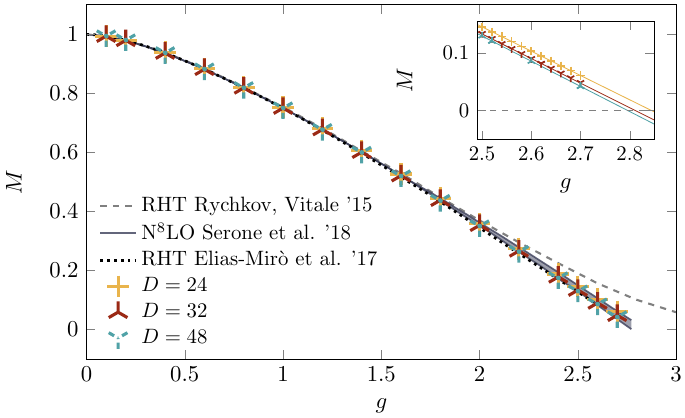}
    \caption{ Results for the mass gap in $\phi^4$-theory as a function of the coupling $g$ obtained from the bootstrap approach. These results are compared with two renormalized Hamiltonian truncation calculations \cite{Elias-Miro:2017xxf,Rychkov:2014eea} (the latter were obtained at finite volume $L=10$ for $m=1$) and Borel-resummed perturbation theory up to order $g^8$ \cite{Serone:2018gjo}. The upper-right inset shows a zoomed-in view close to the critical point, including the linear fits.}     \label{fig:M_vs_g}
\end{figure}

We repeat the analysis for increasing values of the coupling $g$ in Fig.~\ref{fig:M_vs_g} to compare our estimate with renormalized Hamiltonian truncation \cite{Elias-Miro:2017xxf,Rychkov:2014eea} and Borel-resummed perturbation theory \cite{Serone:2018gjo}. Our results for the mass gap agree very well with both approaches across the range of couplings. Remarkably, near the critical point, our bootstrapped estimates are closer to the latest (more precise) RHT and Borel-resummed results than to earlier RHT data \cite{Rychkov:2014eea}. Since our model is in the Ising universality class, renormalization-group arguments imply that the physical mass gap scales as $M \sim C |g-g_c|$ where $C$ a theory-dependent constant~\cite{Rychkov:2014eea}. We thus fit $g_c$ from our $M(g)$ data for $g\in [2.5,2.7]$, and independently for bond dimensions $D=24,32,48$, \textit{i.e.} without finite entanglement scaling \cite{Tagliacozzo:2007rda,Pollmann:2009lnv}. Without further extrapolations, we obtain $g_c = 2.796$ at $D=48$ (at $D=32$, we get $g_c = 2.812$, the difference with $D=48$ provides an order of magnitude of the finite $D$ systematics). This result is already more accurate than existing continuum approaches and is consistent with the best lattice extrapolations \cite{Rychkov:2014eea,Elias-Miro:2017xxf,Serone:2018gjo,Delcamp:2020hzo,Vanhecke:2021noi,Durr:2025hfu}. 


\paragraph{Conclusion} --~~ 
In this letter, we have discussed a central question in nonperturbative quantum field theory: how much information about the dynamics of a relativistic QFT can be extracted directly from an accurate approximate ground state? 

Our results for the mass gap of $\phi^4$ theory at various couplings are in excellent agreement with techniques that directly compute the gap---renormalized Hamiltonian truncation and Borel-resummed perturbation theory---showing that the ground state alone contains precise dynamical information. It will be interesting to compare this approach to tangent-space excitation methods \cite{Vanderstraeten:2019voi,Draxler:2013oek}, if they can be extended from CMPS to RCMPS.

Future applications of the method include the study of multiple one-particle excitations, for instance in the case of two coupled scalar bosons \cite{tiwana2025multifieldrelativisticcontinuousmatrix}; or bound states, such as kink and anti-kink states in the symmetry-broken phase of $\phi^4$-theory~\cite{Serone:2019szm, Gillman:2017uir, Gillman:2017ycq, Rychkov:2015vap}. 

Several improvements are also possible. Our method still relies on a discretization of $\rho$, which could in principle be removed by going to a dual formulation of \eqref{eq:LPreduced} (similarly to what was done in~\cite{Lawrence:2024hjm}). We could also generalize to two-point functions of multiple operators, $\braket{\order_j(x)\order_k(0)}$ simultaneously, which would allow to reconstruct a positive matrix-valued spectral density. In that case, the linear program would become a semi-definite program. It is also natural to extend the approach to higher-point functions, to estimate S-matrix elements, in a spirit similar to recent work combining Hamiltonian truncation and bootstrap techniques~\cite{Karateev:2019ymz,Chen:2021pgx}. Finally, going to higher dimensions is tempting, but precise variational vacua are still not easy to obtain in the continuum, despite recent progress~\cite{eliasmiro2020higherd,eliasmiro2022higherd,tilloy2019ctns,sachar2022ctns_relativistic,roose2025continuumlimitgaugedtensor}. We hope this letter will stimulate their construction.

\begin{acknowledgments}
This project has received funding from the European Research Council (ERC) from the QFT.zip project (grant agreement No. 101040260).
\end{acknowledgments}

\bibliography{references}
\appendix

\section*{End Matter}

\section{Relativistic continuous matrix product states}
\label{app:RCMPS}
Relativistic continuous matrix product states (RCMPS) are a variational class of states designed to approximate ground states of interacting relativistic quantum field theories in $1+1$ dimensions. They are a relativistic extension of continuous matrix product states (CMPS) \cite{Verstraete:2010ft,haegeman2013calculus,haegeman2010qft}, adapted to the natural tensor-factorization of a free bosonic Fock space. 

To define an RCMPS, we first introduce the Fourier transform
\begin{equation}\label{eq:Fourier}
a(x)=\frac{1}{2\pi}\!\int\! \mathrm{d}k \, e^{ikx} a_k ,
\end{equation}
of the normal modes $a_k$ that diagonalize the free (massive) Hamiltonian. These modes are related to the field operator via the standard expansion
\begin{equation} 
\phi(x) = \frac{1}{2\pi}\int_\mathbb{R} \frac{\upd k}{\sqrt{2\omega_{k}}} (a_{k}e^{ikx} + a^{\dagger}_{k}e^{-ikx})~ \, .
\end{equation}
Importantly, $[a(x),a^\dagger(y)] = \delta(x-y)$, and thus the Hilbert space factorizes into a continuous tensor product of Fock space associated to each position. However, commutation was obtained by removing the $1/\sqrt{2\omega_k}$ factor in \eqref{eq:Fourier}, and thus implies that the field $\phi(x)$ and operators $a(x)$ are not local functions of each other, but rather:
\begin{equation}\label{eq:phiRCMPS}
\phi(x) = \int_\mathbb{R}  \mathrm{d}y\, J(x-y) \,[a(y)+a^\dagger(y)]\;,
\end{equation}
with
\begin{align*}
J(x) &= \frac{1}{2\pi}\int\frac{\mathrm{d}k}{\sqrt{2\omega_k}} e^{-ikx}
= \frac{K_{1/4}(|x/m|)}{2^{9/4}\,\sqrt{\pi}\, \Gamma(5/4)\,|x/m|^{1/4}} \; ,
\end{align*}
where $K_\nu$ is a modified Bessel function of the second kind.

The advantage of this operator basis is that the free vacuum $\ket{0}$, annihilated by all $a(x)$, shares the same UV properties as the interacting vacuum $|\Omega \rangle$. In particular, it is expected that the bipartite entanglement entropy of $|\Omega\rangle$ is \emph{finite} when computed in the tensor factorization associated to $a(x)$~\cite{Tilloy:2022kcn,cacciatori2009entropya}. This makes it natural to construct a matrix product state ansatz starting from $\ket{0}$ and using the operators $a(x),a^\dagger(x)$.

On an interval of length $L$ with periodic boundary conditions, an RCMPS is then defined as
\begin{equation}
\label{eq:RCMPS}
\ket{Q,R}
  = \mathrm{tr}\!\left\{
      \mathcal{P}\exp\!\left[ 
      \int_0^L\! \mathrm{d}x\, \left( Q\otimes \mathbf{1} + R\otimes a^\dagger(x) \right)
      \right]
    \right\}\ket{0},
\end{equation}
where $Q$ and $R$ are $D\times D$ complex matrices acting, $\mathcal P\exp$ is the path-ordered exponential, and the trace is taken over the matrix space. The variational parameters are the entries of $Q$ and $R$. Adding a constant multiple of the identity to $Q$ changes only the norm of the state, and we may thus fix it such that the state is normalized. This allows to take the thermodynamic limit $L\rightarrow +\infty$ in all correlation functions.

Computing local observables on a given RCMPS is slightly cumbersome but can be done efficiently~\cite{Tilloy:2021yre,Tilloy:2022kcn}, without ever decompressing the representation into the full Fock space. Ultimately, by solving systems of non-autonomous matrix valued linear differential equations, one can compute the expectation values of local operators at an asymptotic cost equal to that of matrix multiplication \textit{i.e.\!} $\propto D^3$. In particular, since the Hamiltonian density is a sum of local terms, in can be obtained for the same cost. Using backward differentiation techniques, its gradient is also computable at cost $\propto D^3$. 

Minimizing over the space of RCMPS is a non-convex optimization problem, but that is not plagued by the explosion of local minima in practice. Rather, the main difficulty lies in the ill-conditioning of the Hessian, which makes naive gradient descent inefficient. This difficulty is solved with Riemannian optimization techniques~\cite{hauru2021isometric,Tilloy:2022kcn}. In practice, the approximate ground states used in this letter are obtained after a few hundred to a few thousand iterations of the Riemannian \texttt{L-BFGS} algorithm of \texttt{OptimKit.jl}~\cite{OptimKit}.

\section{Estimating integrals of the spectral density}
\label{app:wavefunction_renormalization}
 Reconstruction of fine local features of $\rho(s)$ is challenging, particularly for contributions with small spectral weight, as discussed in the main text. A more robust alternative is to consider global observables, namely integrals of the spectral density over finite energy windows, which probe the cumulative spectral weight in successive regions. These quantities are more stable against small perturbations of $C(x)$ while still encoding nontrivial dynamical information.

Specifically, we replace the objective function \eqref{eq:cost} with integrals of $\rho(s)$ over intervals between multiples of the one-particle mass, $[k\cdot M^2,\, l\cdot M^2]$ (with $M$ estimated, \textit{e.g.}~from the bootstrap procedure outlined above). For instance, in $\phi^4$ theory, the integral from $0$ up to the multiparticle threshold directly yields the wave-function renormalization constant
\be
Z_\phi = \int_0^{s_{\mathrm{th}}} \rho(s)\, \d s \;,
\ee
since the single-particle pole is separated by a finite gap from the onset of the continuum \footnote{In many theories this threshold is $4M^2$; in a $\mathbb{Z}_2$ symmetric model like $\phi^4$ it occurs at $9M^2$.}. 

\begin{figure}[H]
    \centering
    \includegraphics[width=1\linewidth]{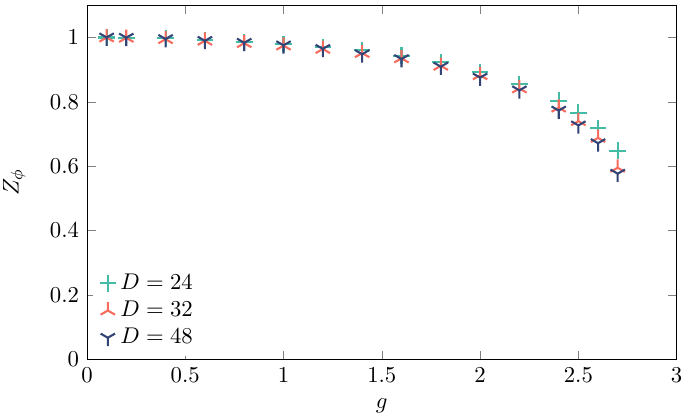}
\caption{Results for the wave-function renormalization $Z_\phi$, \textit{i.e.\!}~the coefficient of the one-particle pole of the renormalized propagator, in $\phi^4$-theory as a function of the coupling $g$ obtained from the linear program \eqref{eq:LPreduced} where we used as objective function $\int_0^{2M^2} \d s\,\rho(s)$, and by bootstrapping the RCMPS slack $\delta C$, until the upper and lower bound coincide. We use $N_c\sim 100$ constraints in the range $x\in [10^{-5},3]$.}\label{fig:Z_vs_g}
\end{figure}

Figure~\ref{fig:Z_vs_g} shows our results for $Z_\phi$ as a function of the coupling $g$, obtained from the linear program \eqref{eq:LPreduced} with objective $\int_0^{2M^2}\rho(s)\, \mathrm{d}s$. We have checked that the results are stable under variations of the upper limit of the integral ($1.5M^2, 2M^2, 2.5M^2$).

Alternatively, $Z_\phi$ can be extracted directly from the bootstrap procedure explained in the main text, which outputs its value at the optimal $M_{\mathrm{opt}}$. Estimates obtained from both methods are highly consistent, differing only by less than $1\%$. 
\vspace{1em}

\end{document}